# Spin orbit torque controlled stochastic oscillators with synchronization and frequency tunability

Punyashloka Debashis[1,2,3], Aman K. Maskay[2], Pramey Upadhyaya[2] and Zhihong Chen[1,2]

[1]Birck Nanotechnology Center, [2]School of Electrical and Computer Engineering,

[3]currently at Intel Corporation, Hillsboro, OR 97124, USA

Purdue University, West Lafayette, USA

*Abstract*—Stochastic oscillators based on emerging nanodevices are attractive because of their ultra-low power requirement and ability to exhibit stochastic resonance, a phenomenon where synchronization to weak input signals is enabled due to ambient noise. In this work, a low barrier nanomagnet based stochastic oscillator is demonstrated, whose output jumps spontaneously between two states by harnessing the ambient thermal noise, requiring no additional power. Utilizing spin orbit torque in a three terminal device configuration, phase synchronization of these oscillators to weak periodic drives of particular frequencies is demonstrated. Experiments are performed to show the tunability of this synchronization frequency by controlling an electrical feedback parameter. The current required for synchronization is more than 8 times smaller than that required for deterministic switching of similar nanomagnetic devices. A model based on Kramers' transition rate in a symmetric double well potential is adopted and dynamical simulations are performed to explain the experimental results.

## Introduction

Oscillators have been an important area of research because of their special properties of synchronization and complex physical interaction that can be leveraged for many applications[1–5]. Research efforts have been targeted towards compact and scalable oscillators based on emerging nanodevices[6–10] because of their low footprint and low power of operation. At the same time, the highly scaled oscillator devices with reduced dimensions are sensitive to thermal fluctuations that affects efficient phase locking and synchronization[11,12]. Recently, there has been emerging research in the area of stochastic oscillators, which utilizes ambient thermal noise to facilitate synchronization. In such oscillators, the state of the device shows self-sustained random fluctuations between two metastable states driven by ambient thermal noise, which determines an average fluctuation rate. A clear advantage of such stochastic oscillators is that no power is required to sustain the fluctuations, which can be converted to periodic oscillations with weak drives, as will be demonstrated in a later section. It has been shown theoretically[13,14] that such oscillators can be synchronized with external periodic excitations through the phenomenon of stochastic resonance[15]. Experimentally, such synchronization has been mostly studied in the field of biological systems[16,17], while very few demonstrations have shown synchronization in silicon CMOS[18,19] or optical devices[20]. Locatelli et al. demonstrated stochastic resonance in a magnetic tunnel junction using spin transfer torque (STT) in a two terminal device geometry[21]. Nevertheless, there is an interest in implementing a three-terminal device, which is one of the necessities to form interconnected networks. Also, electrical frequency tunability, which is a feature of periodic oscillators, has not been shown in the demonstrations of stochastic oscillators mentioned above.

There are three contributions of this work:
- Firstly, it is demonstrated that fluctuations of a low barrier nanomagnet (LBNM) can be synchronized by applying weak sinusoidal currents to a spin orbit torque (SOT) underlayer at a certain frequency, exploiting the phenomenon of stochastic resonance. The required input current amplitude is more than 8 times smaller than that required for deterministic switching of the nanomagnet at zero temperature. Using SOT instead of STT allows a three-terminal geometry for the stochastic oscillator and reduces its power requirement[22].
- Secondly, by utilizing the three-terminal geometry, the output voltage of the device can be fed back to its input through a resistor, which enables the tunability of the average natural frequency of the device. This in turn allows to electrically tune the synchronization condition.
- Finally, the experimental results are corroborated by stochastic dynamical simulations based on Kramers' transition rates in a symmetric double well potential, similar to that introduced by Gammaitoni et al[15].

## Results and discussion

### Low barrier nanomagnet based stochastic oscillator

The average dwell time ($\tau$) of a nanomagnet's magnetization direction is exponentially dependent on the energy barrier ($E_B$) separating its two stable states ("UP" and "DN") through the following expression[23,24]:

$$\tau = \tau_0 exp\left(\frac{E_B}{k_B T}\right) \quad (1)$$

where $\tau_0$ is a material parameter of the magnet called the attempt time, $k_B$ is the Boltzmann constant and $T$ is the ambient temperature. The above expression is valid when $E_B$ > a few $k_B T$, and has to be replaced by a different expression[25] for smaller $E_B$. Nevertheless, when $E_B$ is comparable or smaller than the ambient thermal noise ($k_B T$), the magnetization direction jumps spontaneously between two states, thus forming a stochastic oscillator that requires no additional energy to sustain its fluctuations. For low barrier nanomagnets that satisfy eq. 1, the natural fluctuation frequency ($f_0$) of the stochastic oscillator can be defined as:

$$f_0 = (\tau_{UP} + \tau_{DN})^{-1} = \left(2\tau_0 \exp\left(\frac{E_B}{k_B T}\right)\right)^{-1} \quad (2)$$

Here $\tau_{UP}$ and $\tau_{DN}$ are the average dwell times in the "UP" and the "DN" states and are both equal to $\tau$ in an unbiased nanomagnet. This type of stochastic oscillators has been demonstrated in two-terminal super paramagnetic MTJ devices[26,27].

For a perpendicular magnetic anisotropy (PMA) nanomagnet, the energy barrier $E_B$ is given by the product of its volume ($V$) and the effective anisotropy energy density ($K_{eff}$).

$$E_B = K_{eff} V \quad (3)$$

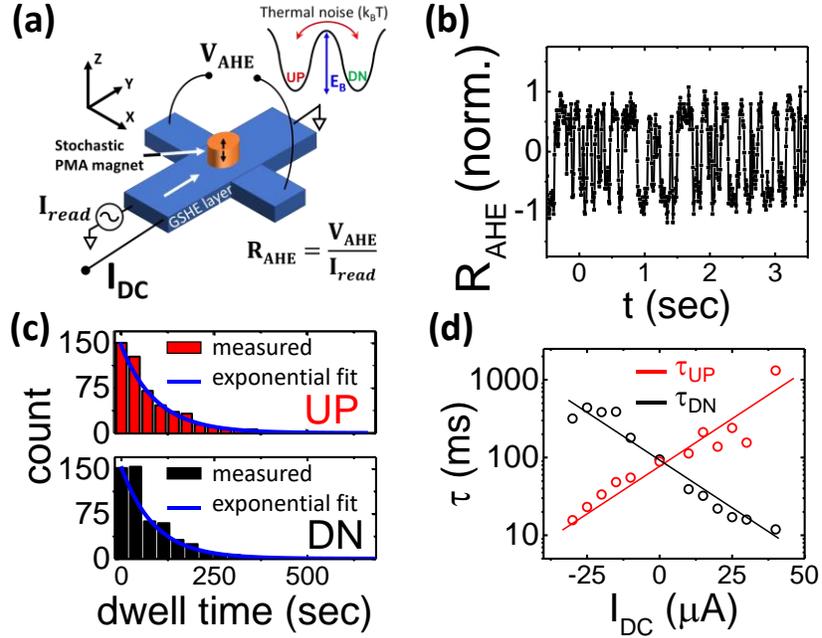

Fig. 1: **Stochastic oscillator device.** (a) Device schematic: a nanomagnet with weak perpendicular anisotropy on a GSHE Hall bar electrode. Its energy barrier is overcome by ambient thermal noise. (b) Device output, $R_{AHE}$ is plotted by normalizing the generated AHE voltage by the magnitude of the read current. The output shows random telegraphic behaviors as the magnetization flips between "UP" and "DN" states. (c) The natural fluctuation frequency ($f_0$) is calculated from the dwell time distribution (d) Dwell times in UP and DN states are tuned by the DC current through the GSHE underlayer.

Here, the effective anisotropy energy density is given by[28]:

$$K_{eff} = \frac{K_i}{t_{FM}} - \frac{M_S^2}{2\mu_0} \qquad (4)$$

where $K_i$ is the interface anisotropy, $t_{FM}$ is the thickness of the ferromagnetic layer, $M_S$ is the saturation magnetization and $\mu_0$ is permeability of free space. The interface anisotropy term $\left(\frac{K_i}{t_{FM}}\right)$ forces the magnetization to point perpendicular to the plane of the film, whereas the demagnetization term $\left(\frac{M_S^2}{2\mu_0}\right)$ favors the magnetization to lie in the film plane. It has been recently demonstrated that by engineering the magnetic layer thickness ($t_{FM}$), the competition between these terms can be carefully balanced to obtain weak PMA stacks[29–31]. This thickness engineering is employed here to obtain a low barrier nanomagnet (LBNM), with $E_B$~18 $k_BT$, showing spontaneous fluctuations in 10s of millisecond time scale. Fig. 1 (a) shows the device schematic, consisting of a perpendicular LBNM island patterned at the center of a 3.5nm thick tantalum (Ta) Hall bar fabricated on a p-Si substrate with 90nm $SiO_2$. The LBMN is a PMA stack of $Co_{60}Fe_{20}B_{20}$(1.3 nm)/MgO(2 nm)/Ta(1 nm). The thickness of 1.3 nm for the $Co_{60}Fe_{20}B_{20}$ layer is carefully chosen to achieve a weak perpendicular anisotropy by partial cancellation of the demagnetization and the interface anisotropy. The diameter of the LBNM is around 70 nm. Due to the combination of weak anisotropy and a small volume, the energy barrier ($E_B$) of the magnet is lowered to ~18 $k_BT$ (as extracted later), resulting in spontaneous fluctuation of the magnetization direction due to ambient thermal noise at room temperature. Anomalous Hall effect (AHE) is utilized to read the magnetization state of the LBNM. A small "read" charge current is passed in the Y direction through the Hall bar to generate a transverse voltage in the X direction, whose sign depends on where the magnetization of the LBNM points in the "UP" (+Z) or "DN" (-Z) direction. Anomalous Hall resistance, $R_{AHE}$, is calculated by dividing the transverse AHE voltage by the read current, and normalized by its maximum value to obtain $R_{AHE}$ (norm.). By taking time resolved measurements, the magnetization can be tracked through $R_{AHE}$ (norm.), as its direction fluctuates randomly. Fig. 1 (b) shows the stochastic oscillations at the output. It is to be noted that current used for this reading scheme is 1 µA, which does not produce any appreciable spin orbit torque on the magnetization.

Fig. 1 (c) shows the histograms of the dwell times of the magnetization in UP and DN states. Both histograms are well fitted by an exponential envelope, suggesting that the underlying mechanism follows a stochastic, Poisson process[21,30]. The average dwell times in the UP and DN state, represented by $\tau_{UP}$ and $\tau_{DN}$ can be obtained from these exponential fits. The natural fluctuation frequency, $f_0$ is then obtained from the sum of $\tau_{UP}$ and $\tau_{DN}$ according to eq. 2. In the presented device, $f_0$ was obtained to be 5 Hz, corresponding to an energy barrier $E_B$ of 18 $k_B T$ at room temperature (assuming $\tau_0$ to be 1 ns[24]). Although the devices shown here have rather low natural fluctuation frequency of a few Hz, they can be made much faster by either reducing the anisotropy or reducing the volume of the LBMN. Owing to the exponential scaling of eq. (1), an LBMN with similar anisotropy as in our experiment, but a diameter of 15 nm would result in a natural fluctuation frequency of ~0.5 GHz, which is attractive for real world applications. Circular in-plane magnets with close to zero anisotropy[32,33] could lead to even faster fluctuations[25,34].

A DC current input to the giant spin Hall effect (GSHE) Ta underlayer produces spin orbit torque on the LBNM and stabilizes one magnetization direction over the other, depending on the polarity of current. A positive DC current stabilizes the UP state and a negative DC current stabilized the DN state, as can be seen by the corresponding increase or decrease in $\tau_{UP}$ and $\tau_{DN}$ in Fig. 1 (d). The reason for this tunability is a slight tilt in the anisotropy axis of the perpendicular LBNM in the X-Z plane, such that a positive current in the +Y direction (which produces spins polarized in the +X direction) favors the magnetization to be in the +Z direction and vice versa. The experimental and theoretical analysis of this tunability have been presented in a previous publication[30]. Heuristically, this effect can be viewed as the DC current producing tilt in the energy landscape on the nanomagnet[21], such that $\tau_{UP,\ DN} =$

$$\tau_0 exp\left(\frac{E_B\left(1\pm\frac{I_{DC}}{I_0}\right)}{k_B T}\right) \tag{5}$$

where the positive and negative sign is for UP and DN state respectively, $I_0$ is the critical switching current of the nanomagnet at zero temperature. By fitting this expression to the semi-log graph of Fig 1 (d), we obtain $E_B$ of 18 $k_B T$ (consistent with previous calculation) and an $I_0$ of 337 µA, which is the critical switching current at zero temperature.

**Synchronization with an external periodic drive**

As described in the previous section, the application of a DC current to the GSHE input of the device can be viewed as producing an effective tilt in the energy landscape towards one of the UP or DN states, depending on the current polarity. When a sinusoidal current is applied to the GHSE input of the device, the energy landscape is periodically tilted towards the UP and the DN states during the positive and negative half cycle of the sinusoid, respectively. When the frequency of this sinusoidal perturbation to the energy landscape matches that of the average transition frequency of the magnetization between the two states, the phenomenon of stochastic resonance[15] takes effect. In this case, the magnetization periodically fluctuates between the two states, in synchronization with the external periodic perturbation. It is to be noted here that the amplitude of the external periodic perturbation is much smaller than that required for deterministic forcing of the magnetization to one state or the other. Fig. 2 (a) shows the measurement configuration with the external sinusoidal current provided to the device input. With the input turned off, the device output shows random fluctuations as seen in the dwell time histogram of its time trace. The histograms are well fitted by an exponential envelope (Fig. 2 (b)) to extract the natural fluctuation frequency ($f_0$) to be 5 Hz for this device, like the one in Fig. 1.

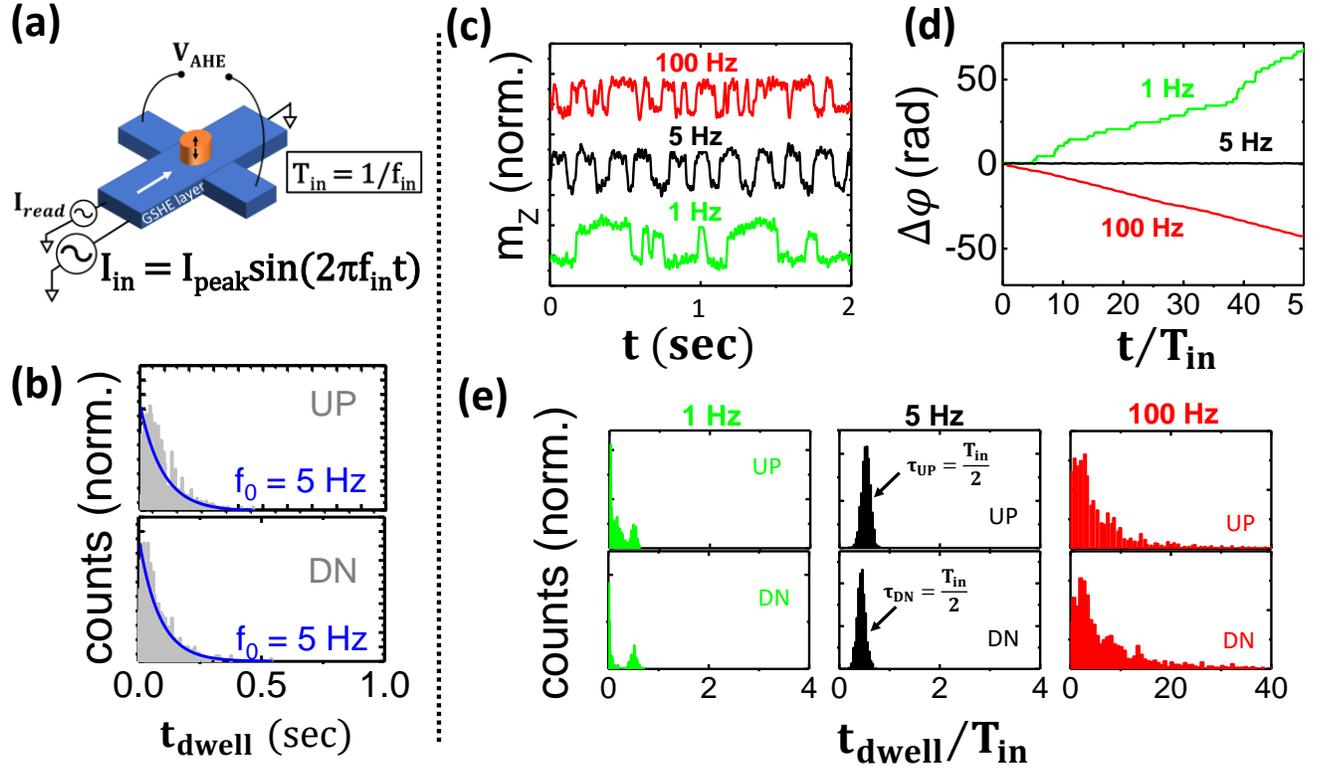

Fig. 2: **Synchronization through stochastic resonance.** (a) Device configuration with a sinusoidal input current with a fixed frequency $f_{in}$ and fixed amplitude ($I_{PP}$ = 42.5 μA). (b) Device behavior with the input turned off. The dwell time histograms shown having exponential envelopes confirm random telegraphic nature of the output. (c) Response of the device to sinusoidal input currents with different frequencies. The output looks periodic when the input frequency $f_{in}$ matches with the natural fluctuation frequency $f_0$ of the device, i.e, 5 Hz. (d) This is quantified by plotting the phase difference between the input and the device output. For the 5 Hz input, the phase difference stays very close to zero (e) This is also reflected in the dwell time histograms. The peak at $t_{dwell}/T_{in}$ = 0.5 for $f_{in}$ = 5Hz shows that the magnetization state stays in one state for half the drive time period before flipping to the other state in sync with the drive. This peak structure is not present for other input frequencies.

Next, a sinusoidal current with amplitude ($I_{peak}$ = 42.5 μA) is provided to the device input and its output is recorded, as illustrated in Fig. 2(a)). This measurement is then repeated for different frequencies ($f_{in}$) of the input sinusoidal current. Fig. 2 (c) shows the output time traces of the device for three different input frequencies, while Fig. 2 (e) plots the dwell time histograms for these time traces. First, for $f_{in}$ = 100 Hz, which is much larger than $f_0$, it can be seen that the device output is random and the corresponding dwell time histogram has an exponential envelope. This suggests that for $f_{in}$= 100 Hz, the device output is unaffected by the input current. Next, when $f_{in}$ is changed to 5 Hz, matching $f_0$, the device output becomes periodic and the dwell time histogram does not have an exponential envelope, rather shows a peak structure around $\tau_{UP} = \tau_{DN} = T_{in}/2$ (where $T_{in}= 1/f_{in}$), as expected for periodic output with the same frequency as that of the input signal. Finally, when $f_{in}$ is further reduced to 1 Hz, the output has a weak dependence on the input and switches sign when the input current is reversed. However, there are a large number of small dwell time transitions or "glitches". This is also evident in the dwell time histogram, where a peak is seen around $T_{in}/2$, but there is a large population of small dwell times, corresponding to the glitches discussed before. In this case, since the input current remains at a particular polarity for longer time than the average dwell time (τ) of the LBNM, the probability of random switching events of the magnetization within one half cycle of the input sinusoid increases.

In order to quantify the synchronization of the device output to the input sinusoidal current, the phase difference ($\Delta\varphi = \varphi_{out} - \varphi_{in}$) between the two signals is evaluated and its evolution as a function of time ($t$) is plotted in Fig. 2(d). Here, the phase $\varphi_{in}$ is determined by the frequency of the input sinusoidal signal according to:

$$\varphi_{in} = 2\pi f_{in} \times t \tag{6}$$

To define the phase of the output stochastic signal, we use linear reconstructed phase using the following expression, similar to that used by Freund et al.[14]:

$$\varphi_{out} = \frac{t-t_k}{t_{k+1}-t_k}\pi + k\pi \tag{7}$$

where $t_k$ is the $k^{th}$ zero crossing of the output signal and $t_k < t < t_{k+1}$

Fig. 2 (d) shows this phase difference for different input frequencies. For $f_{in} = f_0 = 5$ Hz, the output signal of the device becomes periodic with the same frequency as that of the input. Hence, the phase difference stays at zero, showing complete phase synchronization of the input and output signals. For the other two frequencies, the phase difference diverges indefinitely. When $f_{in} < f_0$ the fluctuations at the device output are faster on average than the input frequency, and hence the phase difference between the two signals diverges towards +∞. For $f_{in} > f_0$, the phase difference diverges towards -∞, as can be seen in the figure.

It is to be noted that the amplitude of the input sinusoidal current is much smaller than the critical current ($I_0$) required to cause deterministic switching of the magnetization direction. In the presented device, the width of the Hall bar is 200 nm, energy barrier of the nanomagnet is 18 $k_BT$. Assuming a spin Hall angle of 0.07 for the beta-Tantalum underlayer, the critical switching current at zero temperature is calculated to be 360 µA using the expression given by Liu et al.[35] This value is also consistent with the value of $I_0$ = 337 µA in the previous section extracted directly from Fig. 1 (d). Hence, the amplitude of the input current is more than 8 times smaller compared to $I_0$.

**Tuning the synchronization frequency of the device**

In this section, firstly the tunability of the natural fluctuation frequency of the stochastic oscillator device is demonstrated. Following this, it is shown that stochastic resonance of the same device can be made to occur at different input frequencies by utilizing the above tunability.

When the output magnetization state is amplified and fed back to the GSHE underlayer, the magnetization fluctuation becomes slower or faster, depending on the polarity and strength of the feedback (Fig. 3 (a)). This can be understood by again considering the change to the energy landscape of the LBNM. In the positive feedback configuration, when magnetization is in the UP state, the device output feeds back a positive current to its input, thus tilting the energy barrier favoring the UP state, namely the barrier that needs to be overcome to transition from the UP to the DN state is larger than the barrier for the reverse transition. Similarly, when the magnetization is in the DN state, barrier for going from the DN to the UP state is larger than the barrier for the reverse transition. So, the energy barrier is dynamically modified in a way such that the energy barrier appears to be higher to transition from the occupied state to the other state. This in effect reduces the $f_0$ of the device. The situation is exactly opposite in the case of the negative feedback configuration. In this case, the effective energy barrier to transition from the occupied state to the other state is lowered, hence increasing the $f_0$ of the device. This feedback is achieved in the experiment by amplifying the AHE output voltage and feeding it back through a resistor $R_f$. In this case, the feedback strength can be changed by changing $R_f$. However, for experimental simplicity, the strength and polarity of feedback is tuned by changing the $V_{DD}$ of the amplifier. Fig. 3 (b) shows the time traces of the output fluctuations for different feedback configurations. Fig. 3 (c) shows that $f_0$ can be increased or decreased depending on the feedback strength and polarity. In the y-axis of this plot, $f_0^{int}$ is the intrinsic value of $f_0$ without any feedback. The x-axis shows the feedback voltage ($V_{feedback}$), which is equal to the $V_{DD}$ of the amplifier. This is because due to the high gain of the amplifier, its output voltage always saturates to either $+V_{DD}$ or $-V_{DD}$, which then determines the feedback current through $R_f$.

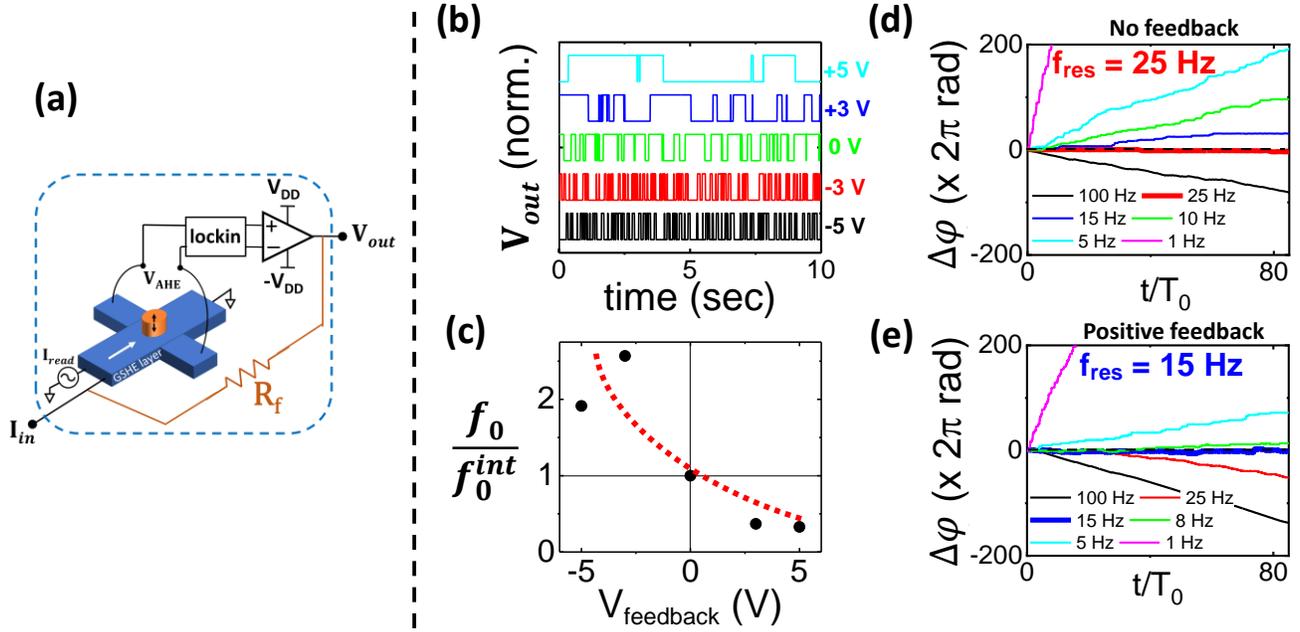

Fig. 3: **Synchronization frequency tunability through feedback.** (a) Feedback configuration. The AHE output voltage is amplified and feedback to the input through the feedback resistor $R_f$. (b) Time traces of the device output for different feedback configurations. The electrical feedback provided by amplifying the output and connecting it to the input through a resistor, $R_{weight}$. The voltage labels for each time trace corresponds to the $V_{DD}$ of the amplifier used. Negative $V_{DD}$ values mean negative feedback. (c) The mean oscillation frequency vs. $V_{feedback}$, which in our experiment equals to $V_{DD}$ of the amplifier, quantifying the effect of the feedback. Here $f_0^{int}$ is the natural fluctuation frequency of the device without any feedback. (d) – (e) show the measurements of synchronization to an external sinusoidal drive with and without feedback. (d) shows the phase difference between the device output and an external sinusoidal input for different frequencies of the input. The synchronization frequency for the device output is 25 Hz for no feedback configuration. (e) shows the results of the same measurement but with an additional positive feedback in place. The synchronization frequency changes from 25Hz to 15 Hz in this positive feedback configuration.

Next, the above feedback method is used to tune the stochastic resonance based synchronization condition with an external sinusoidal input. With no feedback, the device output synchronizes with an input sinusoidal current when $f_{in}$ = 25 Hz, as seen in the phase plots of the device output and the sinusoidal input. Now, by introducing a positive feedback, $f_0$ is reduced and the device output synchronizes with input when $f_{in}$ = 15 Hz (Fig. 3 (e)). Thus, electrically tunable synchronization of the stochastic oscillators is demonstrated.

**Numerical simulations of the dynamics**

The experimental results presented above can be understood phenomenologically by considering the change in Kramers' transition rate[13,35] from UP ($m = +1$) to DN ($m = -1$) state and vice versa as a result of current applied to the GSHE underlayer of the device. The sensitivity of the energy barrier to the applied charge current is captured by the constant $I_0$, the critical switching current at zero temperature as expressed in eq. (5) earlier. We rewrite eq. (5) here in the form of the effective barrier to escape from state $m = \pm 1$ as a function of applied charge current $I_{in}$:

$$E_{B,m} = E_B \left(1 + m\frac{I_{in}}{I_0}\right) = E_B \left(1 \pm \frac{I_{in}}{I_0}\right) \tag{8}$$

And the average dwell times in the UP and DN states modulated by an external current becomes

$$\tau_m = \tau_0 \exp\left(\frac{E_{B,m}}{k_B T}\right) = \tau_k \exp\left(m \frac{I_{in}}{I_0} \frac{E_B}{k_B T}\right) \tag{9}$$

where, $\tau_k = \tau_0 \exp\left(\frac{E_B}{k_B T}\right)$ is the natural average dwell time of the system.

Within this model, the tunability of oscillation frequency automatically emerges as a direct consequence of self-feedback. When a state dependent feedback current $m * \gamma$ is applied, energy barrier and dwell times are adjusted as follows:

$$\tilde{E}_{B,m} = E_B \left(1 + m\frac{I_{in}}{I_0} + \frac{m^2 \gamma}{I_0}\right) = \tilde{E}_B \left(1 + m\frac{I_{in}}{\tilde{I}_0}\right) \tag{10}$$

$$\tilde{\tau}_m = \tilde{\tau}_k \exp\left(m \frac{I_{in}}{I_0} \frac{\widetilde{E_B}}{k_B T}\right) \tag{11}$$

where $\tilde{\tau}_k = \tau_0 \exp\left(\frac{\widetilde{E_B}}{k_B T}\right)$, $\tilde{E}_B = E_B \left(1 + \frac{\gamma}{I_0}\right)$, $\tilde{I}_0 = I_0 + \gamma$, . Equations 10-11 clearly show that a system with self-feedback behaves the same way as a system with a different energy barrier $\tilde{E}_B$ (along with appropriately scaled $\alpha$) and, in turn, with a different natural oscillation frequency $1/2\tilde{\tau}_k$.

Simulations based on extracted parameters and the model described by the Equations 8 and 10 and Kramer' transition rate captures the experimentally observed synchronization as seen in Fig 4. Parameter $I_0$ and energy barrier $E_B$ were extracted from the experimental measurements presented in Fig. 1(d) by calculating the slope and y-intercept respectively of log($\tau$) vs. $I_{DC}$. The parameter $\gamma$ depends on the feedback strength and is extracted from the value of $V_{DD}$ and $R_{feedback}$ in the circuit shown in Fig. 3 (a). Fig. 4a illustrates synchronization phenomena with two distinct measures. When the oscillator is driven by an input with its natural frequency (second row), the dwell time distribution peaks around $T_{in}/2$, which would be the case for a synchronized oscillator. Alternately, the phase difference ($\Delta\varphi = \varphi_{out} - \varphi_{in}$) is bounded within a few cycles, indicating that the input and output are well synchronized. On the other hand, when the drive frequency does not match the average natural frequency of the oscillator (first and third row), dwell time distribution resembles a random Poisson process with no preference to the drive period. Phase difference between the input and output also diverges without bound indicating they are not synchronized.

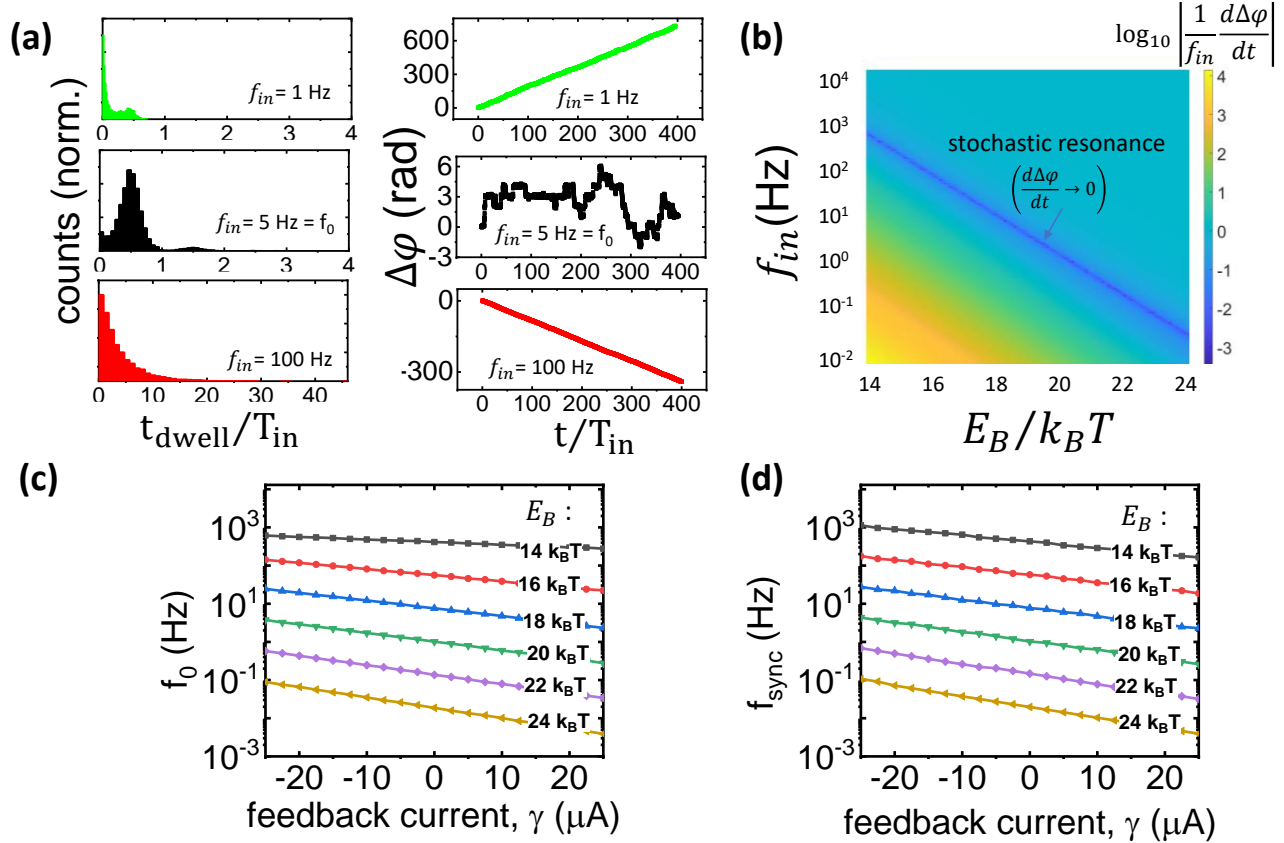

Fig. 4: **Results of numerical simulations.** (a) (left) distribution of normalized dwell times and (right) cumulative deviation in phase over normalized time for 1Hz, 5Hz and 100Hz drive signals. (b) Average deviation in phase velocity of stochastic oscillator normalized to drive frequency. Logarithm is taken to clearly show where synchronization with input drive occurs. (c) Natural fluctuation frequency ($f_0$) vs feedback parameter ($\gamma$) for nanomagnets with different $E_B$. (d) Synchronization frequency (drive frequency that minimizes the cumulative phase deviation), $f_{sync}$ vs $\gamma$ for different $E_B$.

Fig. 4b uses $\frac{d\Delta\phi}{dt}$, the rate of phase divergence, as a measure for synchronization to illustrate how the synchronization frequency changes as energy barrier is varied. Self-feedback leverages this effect to tune the synchronization frequency by controlling the effective energy barrier. Fig. 4c shows the natural fluctuation frequency ($f_0$) changing with different amounts of feedback quantified by parameter $\gamma$. In this case, no input drive is provided, only the output of the device is fed back to the input, which tunes $f_0$. Next, by introducing an external drive in addition to the feedback, the tunability of the synchronization frequency ($f_{sync}$), which is the drive frequency that minimizes the cumulative phase deviation can be studied. It can be seen from Fig. 4(d) that $f_{sync}$ also changes in lock step with the natural fluctuation frequency as the feedback parameter $\gamma$ is changed..

In summary, in this work the phenomenon of stochastic resonance by tuning fluctuations of a weak anisotropy perpendicular nanomagnet using the effect of spin orbit torque is demonstrated in a three terminal device geometry. By applying the output oscillations as feedback to the device input, it was shown that the natural fluctuation frequency and in turn the resonance condition can be tuned electrically. The experimentally observed results were explained by simulations based on Kramers' transition rates in a double well potential. The novel stochastic oscillator device and its unique properties demonstrated in this work present an attractive option for future implementations of energy efficient oscillator-based applications.

## Acknowledgement

This work was supported by the Center for Probabilistic Spin Logic for Low-Energy Boolean and Non-Boolean Computing (CAPSL), one of the Nanoelectronic Computing Research (nCORE) Centers as task 2759.003 and 2759.004, a Semiconductor Research Corporation (SRC) program sponsored by the NSF through CCF 1739635002E.